\documentclass[a4paper]{jpconf} \usepackage{epsfig}
\usepackage{graphicx} \usepackage{amsmath}
\begin{document}
\title{$\Delta G/G$ results from the COMPASS experiment for $Q^2>1$
(GeV/c)$^2$, using high $p_T$ hadrons}

\author{Lu\'{i}s Silva~\footnote{Written on behalf of the COMPASS Collaboration.}}

\address{LIP -- Laborat\'{o}rio de Instrumenta\c{c}\~{a}o Fis\'{i}ca
  Experimental de Part\'{i}culas,\\Av. Elias Garcia, 14 1$^\circ$
  1000-149 Lisboa, Portugal}

\ead{lsilva@lip.pt}

\begin{abstract} 
One of the goals of COMPASS experiment~\cite{compass} is the
determination of the gluon polarisation, $\Delta G/G$, for a deep
understanding of the spin structure of the nucleon.  The gluon
polarisation can be measured via the Photon-Gluon-Fusion (PGF)
process. One of the methods to identify this process is selecting high
$p_T$ hadron pairs in the final state~\cite{Bravar:1997kb}.  The data
used for this analysis were collected by the COMPASS experiment during
the years 2002 to 2006, using a 160 GeV naturally polarised positive
muon beam impinging on a polarised nucleon target.  A new result of
$\Delta G/G$ from high $p_T$ hadron pairs in events with $Q^2>1$
(GeV/c)$^2$ is presented. This result has a better precision due to
the addition of 2006 data and an improved analysis based on a neural
network approach. The gluon polarisation is also presented in three
bins of $x_G$.
\end{abstract}

\section{The Gluon Polarisation and The High $p_T$ Analysis Formalism}
\label{sec:analysis}
The nucleon spin sum rule can be written in a heuristic way as:
$\frac{1}{2}=\frac{1}{2}\Delta \Sigma + \Delta G + L$, where $\Delta
\Sigma$ and $\Delta G$ are, respectively, the quark and gluon
contributions to the nucleon spin and $L$ is the orbital angular
momentum contribution coming from the partons. The purpose of this
analysis is to estimate the gluon polarisation, $\Delta G/G$. The
analysis is performed in two complementary kinematic regions: $Q^2 <1
\ (\mbox{GeV}/c)^2$ (low $Q^2$)~\cite{Ageev:2005pq} and $Q^2 >1 \
(\mbox{GeV}/c)^2$ (high $Q^2$) regions. The present work is mainly
focused on the analysis for high $Q^2$.

Spin-dependent effects can be measured experimentally using the
helicity asymmetry $A^{\rm exp}_{\rm LL}$ defined as
$\frac{\sigma^{\overleftarrow{\Leftarrow}}-\sigma^{\overleftarrow{\Rightarrow}}}{\sigma^{\overleftarrow{\Leftarrow}}+\sigma^{\overleftarrow{\Rightarrow}}}$
where ($\overleftarrow{\Leftarrow}$) and
($\overleftarrow{\Rightarrow}$) refer to the parallel and
anti-parallel spin helicity configuration of the beam lepton
($\leftarrow$) with respect to the target nucleon ($\Leftarrow$ or
$\Rightarrow$). According to the factorisation theorem in DIS, the
(polarised) cross sections can be written as the convolution of the
(polarised) parton distribution function, ($\Delta$)$q_i$, the hard
scattering partonic cross section, ($\Delta$)$\hat{\sigma}$, and the
fragmentation function.

The gluon polarisation is measured directly via the Photon-Gluon
Fusion (PGF) process, which allows to probe the spin of the gluon
inside the nucleon. To tag this process directly in DIS a high $p_T$
hadron pair data sample is used to calculate the helicity
asymmetry. Two other processes compete with the PGF process in leading
order QCD approximation, namely the virtual photo-absorption leading
order process (LP) and the gluon radiation (QCD Compton) process. The
helicity asymmetry for the high $p_T$ hadron pair data sample can thus
be schematically written as:

\begin{equation}
A_{\rm LL}^{2h}(x_{Bj})= R_{\rm PGF} \, a_{\rm LL}^{\rm
PGF}\frac{\Delta G}{G}(x_G) + R_{\rm LP} \, D \, A_1^{\rm LO}(x_{Bj})
+ R_{\rm QCDC} \, a_{\rm LL}^{\rm QCDC} A_1^{\rm LO}(x_C)
\label{eq:allmain}
\end{equation}

The $R_i$ are the fractions of each process, $i$ refers to the
different processes. $a_{\rm LL}^i$ represents the partonic cross
section asymmetries, $\Delta\hat{\sigma}^i/\hat{\sigma}^i$, also known
as analysing power. $D$ is the depolarisation factor, which is the
fraction of the muon beam polarisation transferred to the virtual
photon. The virtual photon asymmetry $A_1^{\rm LO}$ is defined as
$A_1^{\rm LO} \equiv \frac{\sum_i e_i^2 \Delta q_i}{\sum_i e_i^2
q_i}$. This asymmetry $A_1^{\rm LO}$ is estimated using a
parametrisation based on the $A_1$ asymmetry of the inclusive data
~\cite{compassrho}. A similar equation to (\ref{eq:allmain}) can be
written to express the inclusive asymmetry of a data sample, $A_{\rm
LL}^{incl}$.

Using equation~(\ref{eq:allmain}) for the high $p_T$ hadron pair
sample and the above mentioned equation for the inclusive sample the
following expression is obtained:

\begin{align}
  \frac{\Delta G}{G}&(x_G^{av}) = \frac{A_{\rm
      LL}^{2h}(x_{Bj})+A^{corr}}{\lambda}& ~&\text{with :}\qquad
      x_G^{av}=\frac{\alpha_1 x_G - \alpha_2
      x_G'}{\lambda},\label{eq:dgg}\\ 
&\lambda= \alpha_1-\alpha_2~,      
&&A^{corr}= - A_1(x_{Bj})D \frac{R_{\rm LP}}{R_{\rm LP}^{incl}} -
      A_1(x_C) \beta_1 + A_1(x_C')\beta_2,\label{eq:Acorr}\\
      \alpha_1&= a_{\rm LL}^{\rm PGF} R_{\rm PGF} - a_{\rm LL}^{incl,{\rm
      PGF}} R_{\rm LP}\frac{R_{\rm PGF}^{incl}}{R_{\rm LP}^{incl}}~, &\beta_1& = \frac{1}{R_{\rm LP}^{incl}}\bigg[a_{\rm LL}^{\rm
      QCDC} R_{\rm QCDC} -a_{\rm LL}^{incl,{\rm QCDC}} R_{\rm
      QCDC}^{incl}\frac{R_{\rm LP}}{R_{\rm
      LP}^{incl}}\bigg],\label{eq:alphas}\\
      \alpha_2 &= a_{\rm
      LL}^{incl,{\rm PGF}}R_{\rm QCDC}\frac{R_{\rm PGF}^{incl}}{R_{\rm
      LP}^{incl}}\frac{a_{\rm LL}^{\rm QCDC}}{D}~, & \beta_2 &= a_{\rm
      LL}^{incl,{\rm QCDC}}\frac{R_{\rm QCDC}^{incl}}{R_{\rm
      LP}^{incl}}\frac{R_{\rm QCDC}}{R_{\rm LP}^{incl}}\frac{a_{\rm
      LL}^{\rm QCDC}}{D}\label{eq:betas}
\end{align}

Due to the fact that $\Delta G/G$ is present in formula~(\ref{eq:dgg})
at two different $x_G$ (noted $x_G$ and $x_G'$), the extraction of
$\Delta G/G$ requires a definition of the averaged $x^{av}_G$ at which
the measurement is performed.

\section{Data Selection}
\label{sec:data}
Data from years 2002 to 2006 are used. These data were obtained from
polarised muons of 190 GeV/c scattered off a polarised LiD target at
the COMPASS experiment at the CERN SPS. The selected events have an
interaction vertex containing an incoming muon and a scattered
muon. The data are divided into two sets: the high $p_T$ hadron pair
and the inclusive data sample. Both data sets have the $Q^2>1 \
(\mbox{GeV}/c)^2$ kinematic cut applied. Another cut is applied on the
fraction of energy taken by the virtual photon, $y$: $0.1 < y <
0.9$. These cuts are used to select the {\em inclusive sample}.

In the high $p_T$ hadrons data sample, besides the inclusive
selection, at least two outgoing high $p_{T}$ hadrons are
required. These so-called hadron candidates must fulfil the following
requirements: the hadrons with the highest transverse momentum must
have $p_T > 0.7 \ \mbox{GeV/c}$ and the hadron with second highest
transverse momentum, $p_T > 0.4 \ \mbox{GeV/c}$ . This requirement
constitutes the high $p_T$ cut.

\section{The Weighted Method Approach and The Neural Network for $\Delta G/G$
  Extraction}
\label{sec:NN}

The purpose of this analysis is to calculate the gluon polarisation,
$\Delta G/G$, in an event-by-event basis using an optimal weight which
improves the figure of merit. The asymmetry used for the $\Delta G/G$
extraction is related with
the experimental helicity asymmetry, $A^{\rm exp}_{\rm LL}$, described
in section \ref{sec:analysis}, using a weigthing factor $w$.
The correct weight, in this case, should
take into account all the variables that appear in the set of
equations (\ref{eq:Acorr}) to (\ref{eq:betas}), namely: $w=fDP_b
\lambda$, where $f$ is the {\it dilution factor}, the fraction of
polarisable target material, $D$ is the {\it depolarisation factor},
the fraction of muon polarisation is transferred to the virtual
photon, $P_b$ is the muon polarisation and $\lambda$ is defined in equation \ref{eq:Acorr}.

In this analysis it is not possible to tag the events of each involved
process, therefore the process fractions $R_i^j$ and the partonic
asymmetries $a_{\rm LL}^{j,i}$ cannot be directly determined from the
data samples. To estimate or parametrise these quantities a neural
network~\cite{Sulej:2007zz} is used. The neural network is trained by
taking as input samples obtained by Monte Carlo (MC) simulation. The result is a parametrisation which is used for the real data to
provide the values for all these quantities in an event by event
basis.

\section{MC Simulation}
\begin{figure}[h]
\begin{center}
\includegraphics[clip,width=0.16\textwidth]{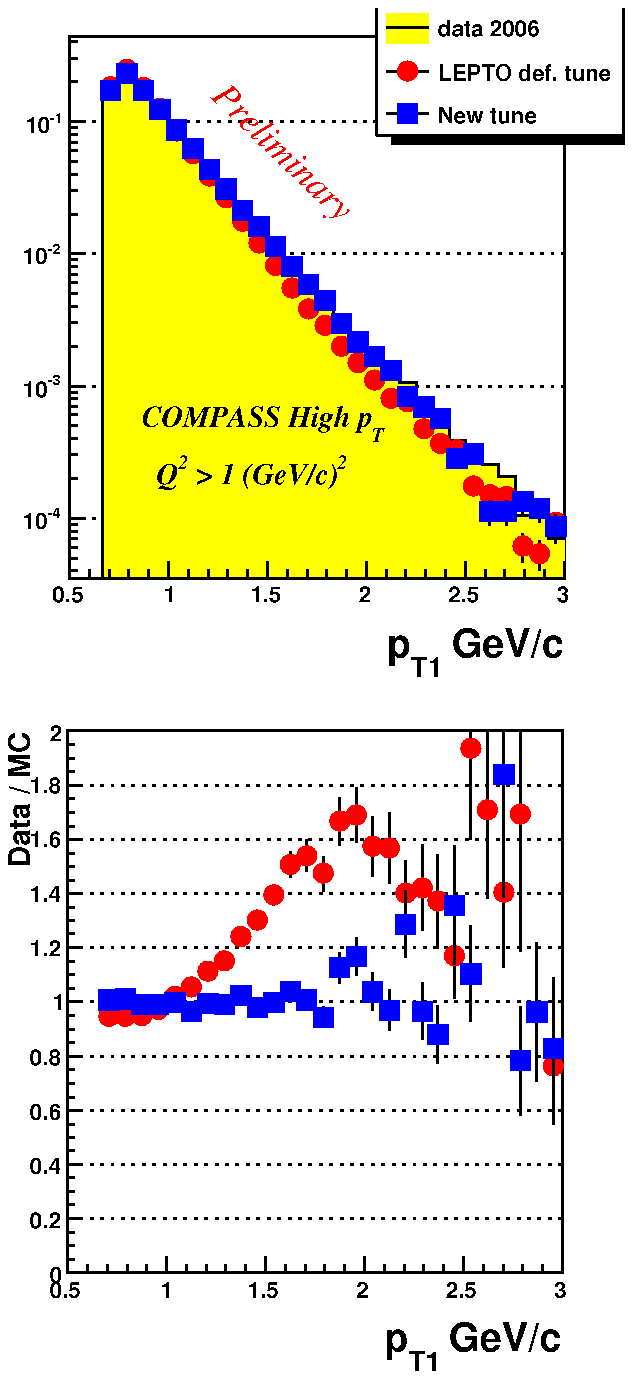}
\includegraphics[clip,width=0.16\textwidth]{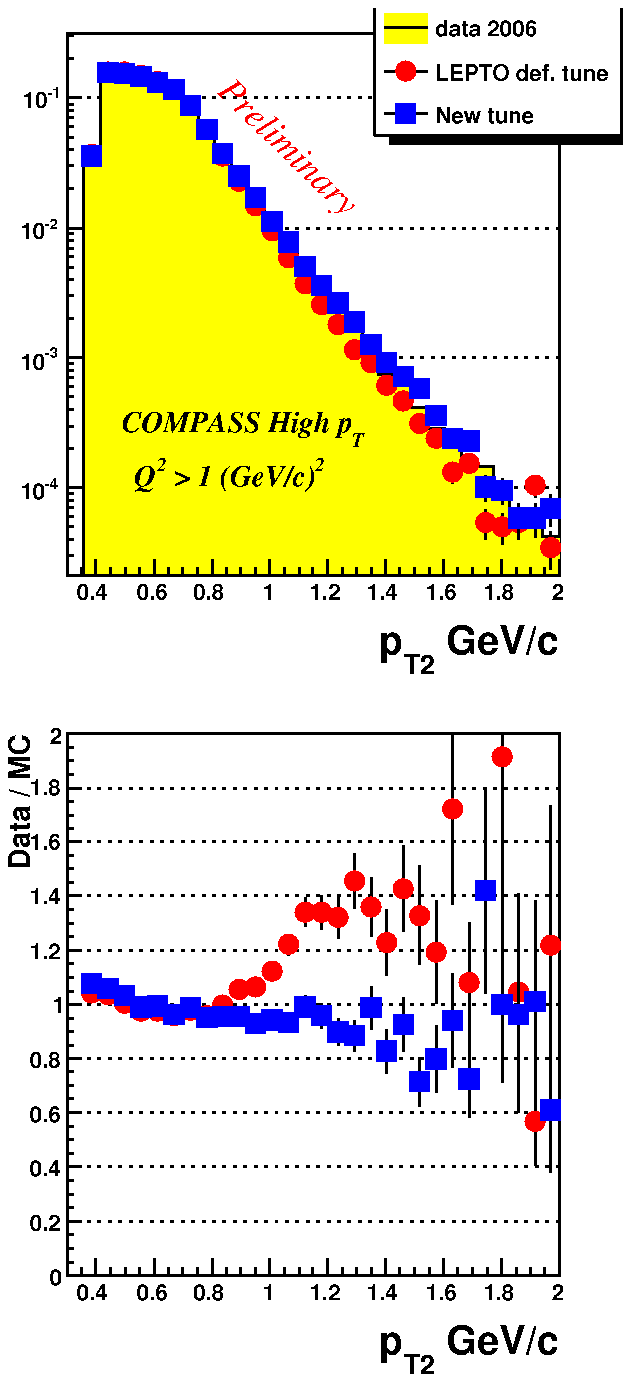}
\includegraphics[clip,width=0.16\textwidth]{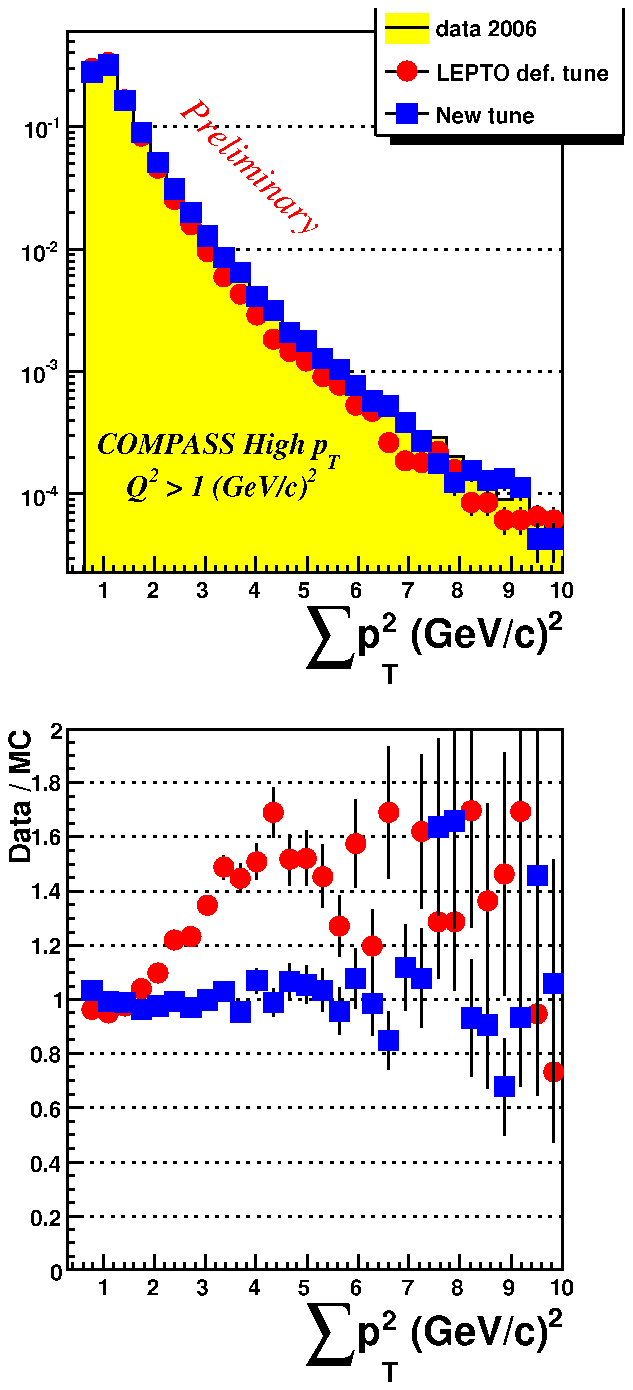}
\includegraphics[clip,width=0.16\textwidth]{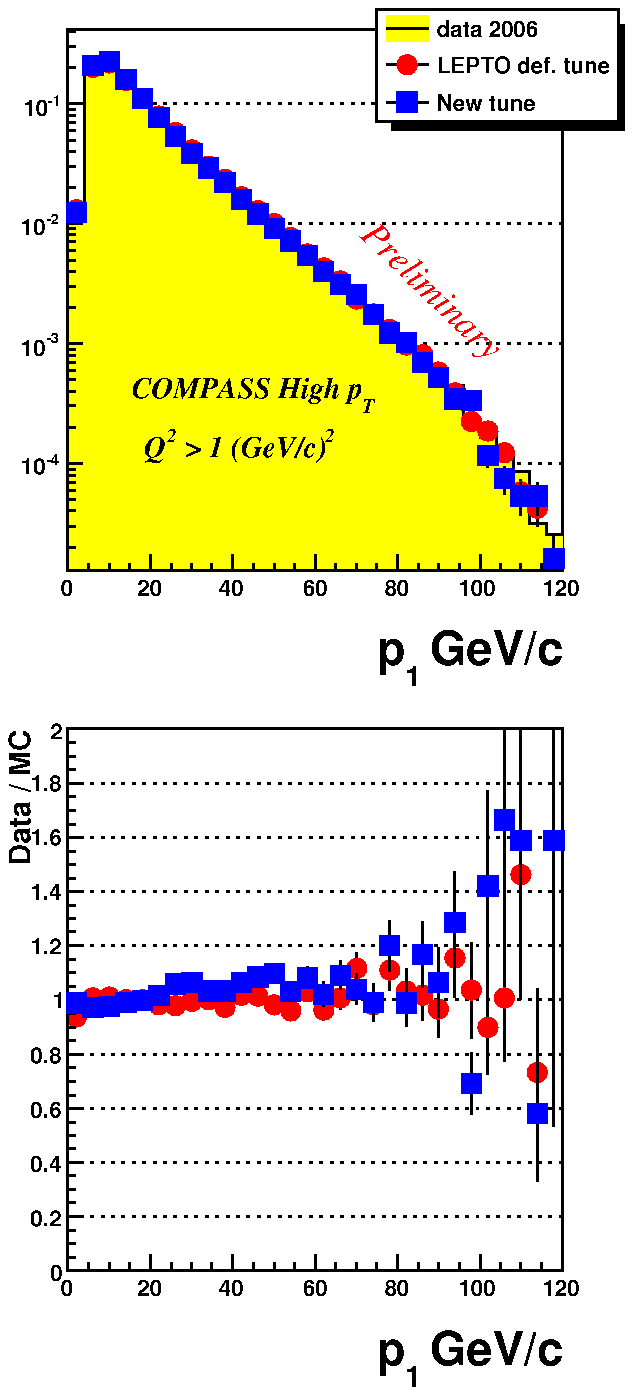}
\includegraphics[clip,width=0.16\textwidth]{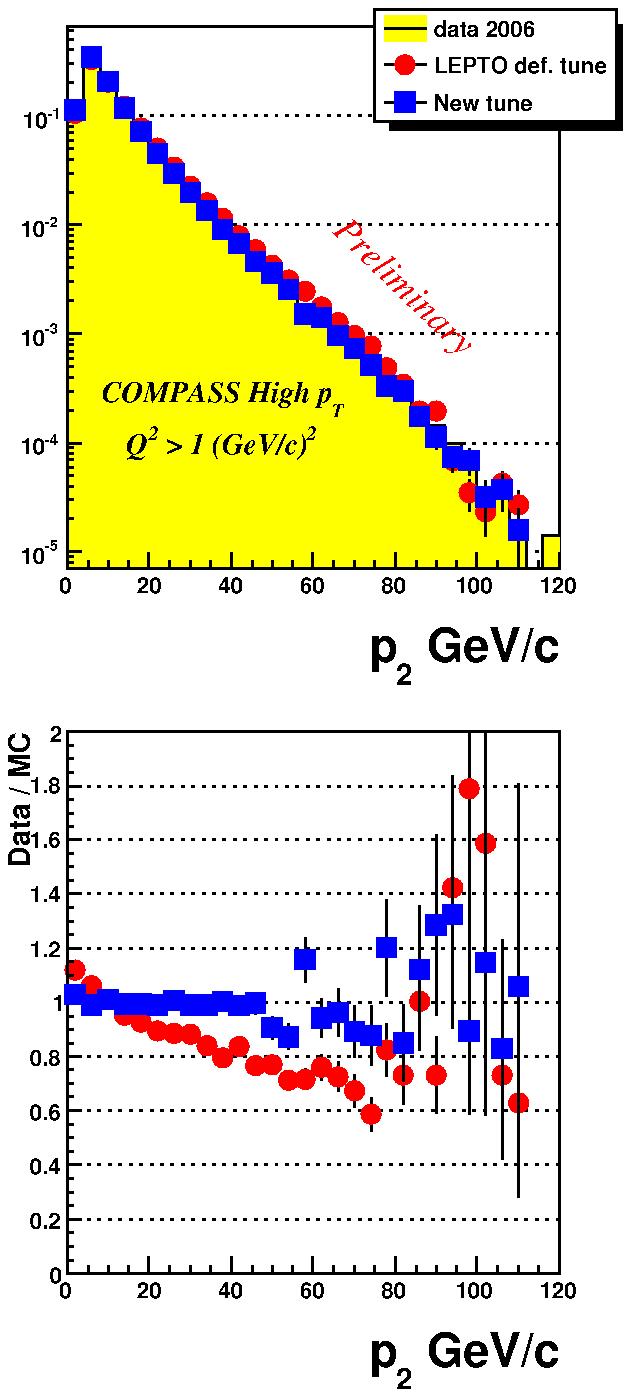}
\includegraphics[clip,width=0.16\textwidth]{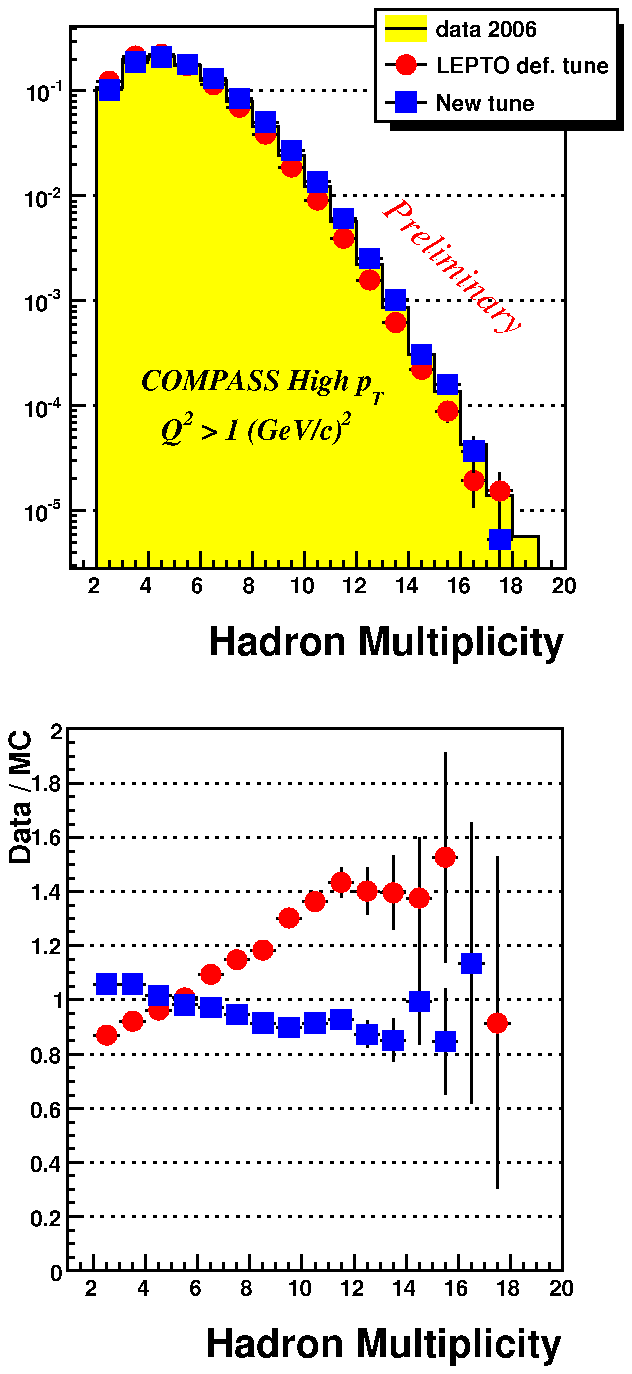}
\caption{Data and MC comparison. Upper plots: (from left to right)
distributions of $p_T$ of the leading hadron, $p_T$ of the sub-leading
hadron, sum of the leading and sub-leading hadron $p_T^2$, $p$ of the
leading hadron, $p$ of the sub-leading hadron and hadron
multiplicity. Lower plots: respective data over MC ratio.}
\vspace{-10pt}
\label{fig:p}
\end{center}
\end{figure}

The event simulation is one of the issues of major importance for this
analysis since several parameters for the $\Delta G/G$ extraction are
taken from the simulation. Thus a strong effort was made to achieve a
simulation very close to the real data. The MC production comprises
three steps: first the events are generated, then the particles pass
through a simulated spectrometer using a program based on GEANT
3~\cite{geant} and finally the events are reconstructed using the same
procedure applied to real data. For the first step the LEPTO 6.5~\cite{Ingelman:1996mq} DIS event
generator is used together with a leading order parametrisation of the
parton distributions. The fragmentation is based on the Lund string
model ~\cite{Andersson:89} implemented in
JETSET~\cite{Sjostrand:1985ys}. Figure~\ref{fig:p} illustrates the quality of the simulation obtained at the end.

The next step was the tuning of the event generator. The Parton
Distribution Function (PDF) set used in this analysis is
MSTW2008LO~\cite{Martin:2009iq}. For the calculation of $F_L$ in LEPTO the ratio
$R(x,Q^2)=\frac{\sigma_L}{\sigma_R}$ is needed. Here the
parametrisation of ref.~\cite{Ingelman:1996mq} was used. To improve the description of the hadrons, higher order QCD
corrections are partially simulated by including gluon radiation in
the initial and final states ({\it parton shower} --
PS)~\cite{Ingelman:1996mq}. An extensive and careful {\it tuning} of
the Lund string fragmentation function parameters and of the hadron
intrinsic transverse momentum parameters was performed.

The tuning was done first on the fragmentation function parameters, in
this case of the the Lund string function: $f(z)\propto
\frac{1}{z}(1-z)^a exp(-\frac{b \cdot m_{\perp}^2}{z})$ where
$m_\perp^2 = m^2 + p_\perp^2$; the parameters are $a$ and $b$ which
are controlled by the JETSET parameters \texttt{PARJ(41)} and
\texttt{PARJ(42)}.

The second step was the tuning of the hadron intrinsic transverse
momentum model, described in~\cite{Sjostrand:1985ys}. The parameters
for this model are {\tt PARJ(21)}, {\tt PARJ(23)} and {\tt PARJ(24)},
which correspond, respectively, to the {\it sigma} of the main
Gaussian, to the {\it sigma} and to the {\it amplitude} fractions for
the second Gaussian. The result of all these improvements is the
COMPASS tuning.

In figure~\ref{fig:p} the kinematic distributions
and some hadronic distributions are shown, details can be found in the
figures caption. In these figures, two MC simulations with different
tunings are shown: the LEPTO default and the COMPASS tuning.  In
general the COMPASS tuning describes better the data, particularly for
the hadronic variables $p_T$ and $\sum p_T^2$ in which a remarkable
agreement between MC and data is presented in the figure. The same
applies for the hadron multiplicity (rightmost in figure~\ref{fig:p}). In table~\ref{tb:tune} the values of the modified parameters are
shown.

\begin{table}[h]
%\vspace{-25pt}
\caption{\label{tb:tune}Modified JETSET MC parameters.}
\begin{center}
\begin{tabular}{llllll}
\br
Tuning&\texttt{PARJ(21)}&\texttt{PARJ(23)}&\texttt{PARJ(24)}&\texttt{PARJ(41)}&\texttt{PARJ(42)}\\
\mr LEPTO Default&0.36&0.01&2.0&0.3&0.58\\
COMPASS&0.34&0.04&2.8&0.025&0.075\\ \br
\end{tabular}
\end{center}
\vspace{-25pt}
\end{table}

\section{Systematic Studies}
The total systematic error is $\delta(\Delta G/G) = 0.063$. The
contributions come from several sources which were studied in
detail. For the neural network, the dependence on the internal
structure was taken into account. For the MC several samples with
different configurations (tuning, PDF, with and without PS, with and
without $F_L$) were studied. An extensive study was performed
searching for false asymmetries. The dependence of the asymmetry
$A^d_1$ using different parametrisations was studied. For equation
(\ref{eq:Acorr}) two approximations were used for $x^\prime_C$: one
with $x^\prime_C= const \cdot x_C$, and another $x^\prime_C$ obtained
from a second iteration in the neural network of $x_C$ given as
input. Also the uncertainty for the input variables related to the
polarisation states of the beam and target: $P_b$, $P_t$ and $f$, were
taken into account. The major contributions come from the MC,
$\delta(\Delta G/G)_{MC} = 0.045$, and the $\Delta G/G$ formula in
equation (\ref{eq:Acorr}), $\delta(\Delta G/G)_\text{formula} =
0.035$~.

\section{Results}
\begin{figure}[h]
\begin{center}
  \includegraphics[width=1\textwidth]{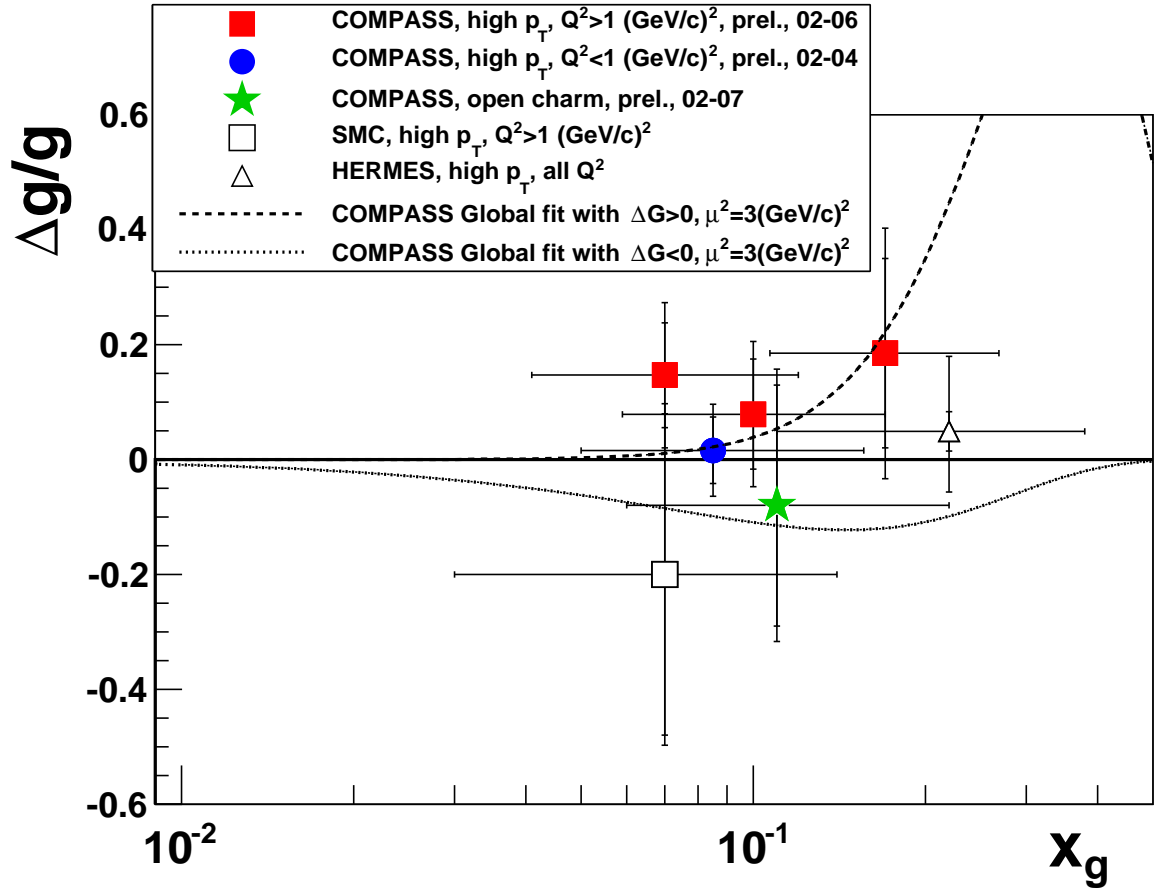}
\end{center}
\caption{\label{fig:dgg} $\Delta G/G$ Results from
  COMPASS~\cite{Ageev:2005pq}, SMC~\cite{Adeva:2004dh},
  HERMES~\cite{Airapetian:1999ib} experiments. Also NLO QCD fits are
  shown~\cite{comp.del_sigma}. }
\end{figure}

The gluon polarisation, $\Delta G/G$, is calculated using equations
(\ref{eq:dgg}) to (\ref{eq:betas}). The result is $\Delta G/G =
0.125\pm 0.060\pm 0.063$ calculated at $x^{av}_G=
0.09_{-0.04}^{+0.08}$~. In order to investigate $x_G$ dependence of
$\Delta G/G$ the data is divided into three bins of the parametrised
$x_G$ variable, {\it i.e.} divided into three independent
samples. These results are given in table \ref{tb:dgg} and also
presented in figure~\ref{fig:dgg}. In the same figure, other results
from the COMPASS collaboration are also depicted, together with
results from HERMES and SMC experiments.

\begin{table}[h]
%\vspace{-25pt}
\caption{\label{tb:dgg}Gluon polarisation results in bins of $x_G$.}
\begin{center}
\begin{tabular}{llll}
\br &$1^{st}$ Bin&$2^{nd}$ Bin&$3^{rd}$ Bin\\ \mr $\Delta
G/G$&$0.147\pm 0.091\pm 0.088$&$0.079\pm 0.096\pm 0.082$&$0.185\pm
0.165\pm 0.143$\\
$x_G^{av}$&$0.07_{-0.03}^{+0.05}$&$0.10_{-0.04}^{+0.07}$&$0.17_{-0.06}^{+0.10}$\\
\br
\end{tabular}
\end{center}
\vspace{-25pt}
\end{table}

\end{document}